\documentclass{optica-article}
\usepackage{siunitx}
\usepackage{xcolor}
\journal{opticajournal} 
\usepackage{float}
\articletype{Research Article}

\begin{document}

\title{Enhancing  broadband second harmonic generation in a thin film lithium niobate racetrack resonator with tunable-coupling}

\author{Olivia Hefti,\authormark{1,2,*} 
Jean-Etienne Tremblay,\authormark{1,$\dagger$} 
Andrea Volpini,\authormark{1}
Jannis Holzer, \authormark{1} 
Alberto Della Torre,\authormark{1}
Homa Zarebidaki,\authormark{1}
Charles Caër,\authormark{1}
Hamed Sattari,\authormark{1,$\ddagger$}
Camille-Sophie Brès,\authormark{2}
 and Davide Grassani,\authormark{1}}

\address{\authormark{1}Centre suisse d'électronique et de microtechnique (CSEM), 2000 Neuchâtel, Switzerland\\
\authormark{2}Photonic Systems Laboratory, École Polytechnique Fédérale de Lausanne, 1015 Lausanne, Switzerland\\}
\email{\authormark{*}olivia.hefti@csem.ch\\} 
\email{\authormark{$\dagger$}Now at: {Enlightra, 1020 Renens, Switzerland}\\}
\email{\authormark{$\ddagger$}Now at: CCRAFT, 2000 Neuchâtel, Switzerland\\}

\begin{abstract*} 
Second harmonic generation in thin film periodically poled lithium niobate (PPLN) is constrained by an efficiency–bandwidth trade-off and fabrication-sensitive scaling. We demonstrate a racetrack resonator incorporating a short PPLN section and a tunable Mach–Zehnder interferometer coupler that enables in situ control of the coupling condition, compensating fabrication tolerances and stabilizing operation near critical coupling. The telecom pump is resonantly enhanced, while the near-infrared second harmonic is generated in single pass, eliminating dual-resonance requirements. The device achieves a 35 times efficiency enhancement over a non-resonant structure while maintaining a $\sim$7 nm bandwidth. This architecture provides a robust platform for broadband integrated frequency doubling.
\end{abstract*}

\section{Introduction}
A long-standing challenge in the design of optical frequency doublers is to achieve high conversion efficiency while maintaining a broad operating bandwidth. In second harmonic generation (SHG), a fundamental trade-off arises because the second harmonic (SH) power scales quadratically with the interaction length ($P_{\mathrm{SH}} \propto L^2$), whereas the phase matching bandwidth scales inversely with the length ($\propto L^{-1}$). Increasing the waveguide length therefore enhances the peak SH power but simultaneously narrows the bandwidth. In practice, extending the device length is not always beneficial as longer waveguides are more susceptible to fabrication-induced inhomogeneities, such as film thickness and etch-depth variations, which degrade phase matching, broaden the spectral response, and reduce the expected peak efficiency \cite{chen2024adapted}. As a result, simply increasing $L$ does not necessarily translate into improved device performance. To overcome this limitation, dispersion-engineering approaches have demonstrated that minimizing the group-velocity mismatch between the pump and the second harmonic can significantly broaden the SHG bandwidth without sacrificing conversion efficiency \cite{wang2018ultrahigh}, particularly in pulsed regimes. An alternative strategy consists of employing resonant structures, which exploit intracavity field enhancement to boost conversion efficiency for a fixed interaction length, thereby preserving a broader envelope bandwidth in the nonlinear medium. 

Three-wave mixing processes, such as optical parametric oscillation (OPO) and spontaneous parametric down-conversion (SPDC), involve three distinct wavelengths and are termed triply resonant when all are resonant. In SHG, the structure is doubly resonant when both the pump and second harmonic resonate, in contrast to singly resonant configurations where only the pump is resonant. Triply (doubly for SHG) resonant structures have been proposed to increase the nonlinear conversion efficiency in Z-cut \cite{lu2019periodically,lu2021ultralow} and X-cut \cite{chen2019ultra,mckenna2022ultra,bruch2019chip,hwang2024spontaneous,briggs2024precise} thin film lithium niobate (TFLN) waveguides, achieving efficiencies as high as $\SI{250000}{\percent \per \watt}$, as well as in other photonic integrated platforms, such as SiN \cite{lu2021efficient,nitiss2022optically}, AlN \cite{bruch201817}, GaP \cite{lake2016efficient} and SiC \cite{lukin20204h}. Despite their impressive efficiency, triply (doubly for SHG) resonant structures present several challenges. First, designing a coupler that operates efficiently at widely separated wavelengths requires a compromise on the coupling conditions, as waveguide-to-ring coupling (cross-coupling) is generally stronger at longer wavelengths. This issue is particularly pronounced when quasi-phase matching enables SHG on the more tightly confined fundamental mode. Second, maintaining triply (doubly for SHG) resonant condition requires aligning resonances at widely separated wavelengths \cite{briggs2024precise}, which is challenging due to dispersion. Third, photorefractive and thermo-optic effects induce wavelength-dependent resonance shifts, complicating stable operation.

Consequently, although triply (doubly for SHG) resonant architectures can deliver very high efficiencies, they often offer limited flexibility and tunability. To overcome these limitations, a triply resonant TFLN racetrack with a tunable Mach-Zehnder interferometer (MZI) coupler has been numerically proposed \cite{kundu2025periodically}. This design enables quasi-independent control of the coupling conditions at telecom and near-visible wavelengths by exploiting the thermo-optic mismatch in the shared MZI heater, to support efficient quantum light generation. The influence of tunable coupling on the coincidence counts, the coincidence-to-accidental ratio (CAR), and the photon-pair generation bandwidth in spontaneous four-wave mixing (SFWM) has been experimentally investigated in a similar SiN design \cite{heine2026tunable}. 
In addition, linearly uncoupled but nonlinearly coupled resonators have shown the ability to individually tune the pump and second harmonic resonances, while optimizing the couplers separately at each wavelength \cite{clementi2025ultrabroadband,stefano2024broadband}. While highly effective in SiN platforms, achieving efficient cross-coupling at near-visible wavelengths is challenging to implement in TFLN, due to the non-vertical sidewalls typically produced in TFLN waveguides \cite{desiatov2019ultra}.

Beyond the advances in the design of triply (doubly for SHG)  resonant cavities, doubly (singly for SHG) resonant designs, where only the longer wavelengths resonate, are increasingly used to improve nonlinear efficiency in PPLN-based TFLN devices. Some approaches use dichroic couplers (directional or adiabatic couplers) \cite{ledezma2023octave,roy2023visible} for wavelength selectivity to enable doubly resonance operation. Such doubly resonant structures have already been used in several demonstrations, such as OPOs for near-IR to mid-IR frequency combs, visible frequency comb, tunable infrared parametric oscillators, and for the generation of single-mode squeezed-light \cite{roy2023visible,ledezma2023octave,park2024single}.
However, achieving the targeted coupling conditions at the desired wavelengths remains a challenge because of fabrication imperfections. Here, we design, fabricate, and experimentally characterize a singly resonant structure in which the PPLN is embedded in one arm of a MZI. This architecture offers three key advantages. First, it preserves the simplicity and operational stability of a singly resonant configuration. Second, it enables tunable coupling of the pump, allowing compensation for fabrication imperfections and dynamic reconfiguration of the device for different operating regimes (overcoupling, undercoupling or critically coupling). Third, by integrating the PPLN directly within the coupler, it eliminates the need for an additional dichroic element to efficiently extract the second harmonic, thereby avoiding the excess losses that such a component might introduce.

\begin{figure*}[ht]
  \centering
\includegraphics[width=14 cm]{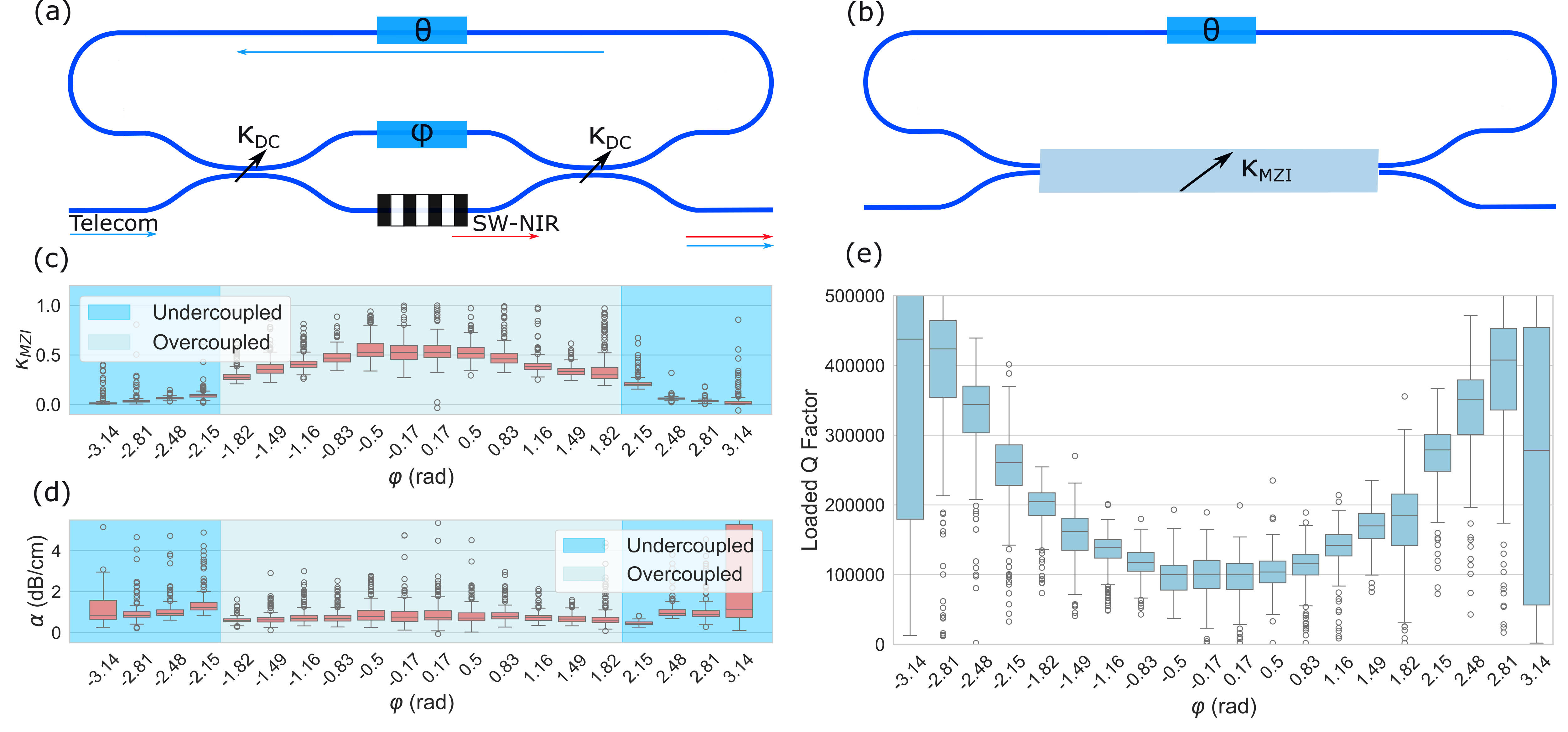}
\caption{(a) Singly resonant racetrack resonator incorporating a MZI coupler with a PPLN section in one arm. (b) Schematic illustrating the effective MZI coupling coefficient $\kappa_{\mathrm{MZI}}$. (c)-(e) Box plots of the extracted (c) $\kappa_{\mathrm{MZI}}$, (d) propagation loss $\alpha$, and (e) loaded quality factor, obtained from Lorentzian fits to transmission spectra measured at different MZI phases $\varphi$. Font colors indicate the under- and over-coupled regimes.}
  \label{fig:pannel1}
\end{figure*}

\section{Results}
\subsection{Racetrack design}
Our singly resonant racetrack for frequency conversion is illustrated in Fig.\ref{fig:pannel1} (a). The coupler between the bus waveguide and the racetrack is a MZI. It is made of two directional couplers (DC) with power coupling coefficient $\kappa_{\mathrm{DC}}$, of a thermal heater in the upper arm, to control the phase $\varphi$, and of a PPLN in the lower arm to perform frequency conversion. As indicated by the arrows, long telecom wavelengths (in blue) are resonant, while short-wavelengths near-infrared (SW-NIR) are single-pass (in red). In the case of SHG, the pump is resonant, while the SH generated in the PPLN is not coupled by the directional coupler and continues straight to the output. A second thermal heater, located in the racetrack, controls the phase $\theta$ and tunes the resonant wavelength. Fig.\ref{fig:pannel1} (b) shows a scheme in which the MZI is considered as a black-box with an effective power coupling coefficient $\kappa_{\mathrm{MZI}}$. Notably, $\kappa_{\mathrm{MZI}}$ can be measured experimentally by fitting the transmitted resonances with Lorentzian functions while sweeping the laser wavelength, following the standard approach used for waveguides coupled to a ring resonator with a point coupler \cite{yariv2000universal}. $\kappa_{\mathrm{MZI}}$ represents the fraction of optical power coupled per round trip from the bus waveguide into the ring. In coupled-mode theory, the external coupling rate $\kappa_{\mathrm{ex}}$ describes how fast energy decays from the cavity into the waveguide in the time-domain. These quantities are related with the relation $\kappa_{\mathrm{ex}} = 2\pi\frac{\kappa_{\mathrm{MZI}}}{T_\mathrm{rt}}$, where $T_\mathrm{rt}$ is the round-trip time of the resonator and $\kappa_{\mathrm{ex}}$ is expressed in $\SI{}{\radian \per \second}$, while $\kappa_{\mathrm{MZI}}$ is unitless. The enhancement of the input pump power, $P_{\mathrm{p,in}}$, by the racetrack, at resonance, is given by the pump power enhancement, $\mathrm{PE}=\frac{P_\mathrm{p}}{P_{\mathrm{p,in}}}$, where $P_\mathrm{p}$ is the pump power within the resonator. Note that there are two resonating paths in our design, as the light can resonate either in the upper or in the lower arm of the MZI coupler.

\subsection{Racetrack fabrication}
The devices are fabricated at wafer scale in a standard foundry process at CSEM \cite{csem_foundry} using commercially available TFLN on insulator wafers. The wafer stack consists of a \SI{600}{\nano\meter}-thick monocrystalline X-cut LN layer on a \SI{4.7}{\micro\meter} buried thermal oxide (BOX). Rounded-tip electrode fingers for electric-field poling are first patterned by electron-beam lithography (EBL), followed by the deposition of a Cr/Au/Cr metal stack and lift-off. The electrodes have a period of approximately \SI{4}{\micro\meter}, a 50\% duty cycle, and a gap of \SI{8}{\micro\meter}. Wafer-level poling is performed on a wafer prober using a waveform generator (Siglent SDG1032X) and a high-voltage amplifier (TREK 2220). After poling, the metal electrodes are removed. Waveguides are then defined by EBL and etched into the LN layer using argon-based dry etching. A second Cr/Au/Cr metal deposition is performed to pattern the integrated heaters via EBL, evaporation, and lift-off. Finally, the waveguides are cladded with SiO$_2$, and the chips are singulated. 

\subsection{Linear characterization}
The devices are characterized using the following experimental setup. Light from a tunable continuous-wave laser (Toptica CTL 1550) is coupled into the chip via a polarization-maintaining lensed fiber (OZ Optics) on the quasi-TE$_{00}$ mode of the waveguide. The chip output is collected and collimated with a lens. A dichroic mirror separates the SH signal from the fundamental pump, which are then directed to two photodetectors (PDA36A-EC and PDA10C-EC) connected to an oscilloscope. The pump and SH are monitored on the oscilloscope. Electrical power is supplied to the integrated heaters using DC probes connected to a voltage source.

Figs.\ref{fig:pannel1} (c) and (d) provide the experimentally extracted power coupling coefficient between the bus waveguide and the racetrack, $\kappa_{\mathrm{MZI}}$, and the propagation loss (including losses in the directional couplers), $\alpha$, for different phase $\varphi$ controlled by tuning the heater power. As expected, the propagation loss is independent of the heater power, yielding an average value of $\alpha = \SI{0.78(0.18)}{\decibel\per\centi\meter}$. In contrast, $\kappa_{\mathrm{MZI}}$ is tunable from a low value of 0.013, corresponding to the undercoupled regime, to a maximum value of $\kappa_{\mathrm{MZI,max}} = 0.53$. The coupling regime at a given voltage is identified by the extinction ratio, which is maximized at critical coupling and delineates the transition between the under- and overcoupled regimes. Fig.\ref{fig:pannel1} (e) shows the measured loaded quality factor (Q) of the pump as a function of $\varphi$, tuned by changing the heater power. The high error bars in the first and last points are due to the fact that, being undercoupled, the peak of resonances are very shallow and can hardly be fitted to extract the Q factor.

\subsection{Resonant field enhancement simulations}
From the experimentally determined $\kappa_{\mathrm{MZI,max}}$ and $\alpha$, we infer the power coupling coefficient of the directional couplers, $\kappa_{\mathrm{DC}}$, assuming they are identical, as per the design. Light propagation within the structure is simulated using SAX, a Python-based frequency-domain solver that models photonic components and integrated circuits using scattering-matrix (S-matrix) formalism. By evaluating the electric field distribution across a range of wavelengths and spatial positions within the device, resonance peaks can be identified and analyzed to quantify field enhancement and extract the quality factor. Key parameters, including $\alpha$, $\kappa_{\mathrm{DC}}$, and $\varphi$ are systematically varied to investigate their impact on the resonant response. To validate the model, we derive an analytical expression for the pump power in the lower arm of the MZI. The theoretical results show excellent agreement with the model, and are presented in the supplementary material (Fig. S2). Fig.\ref{fig:pannel2_a} (a) shows SAX simulations of the MZI coupler obtained by sweeping $\kappa_{\mathrm{DC}}$ and the phase $\varphi$, with the propagation loss fixed to $\alpha = \SI{0.8}{\decibel\per\centi\meter}$. The red line corresponds to the experimentally maximum measured value $\kappa_{\mathrm{MZI,max}} = 0.53$.
\begin{figure}[H]
\centering\includegraphics[width=\linewidth]{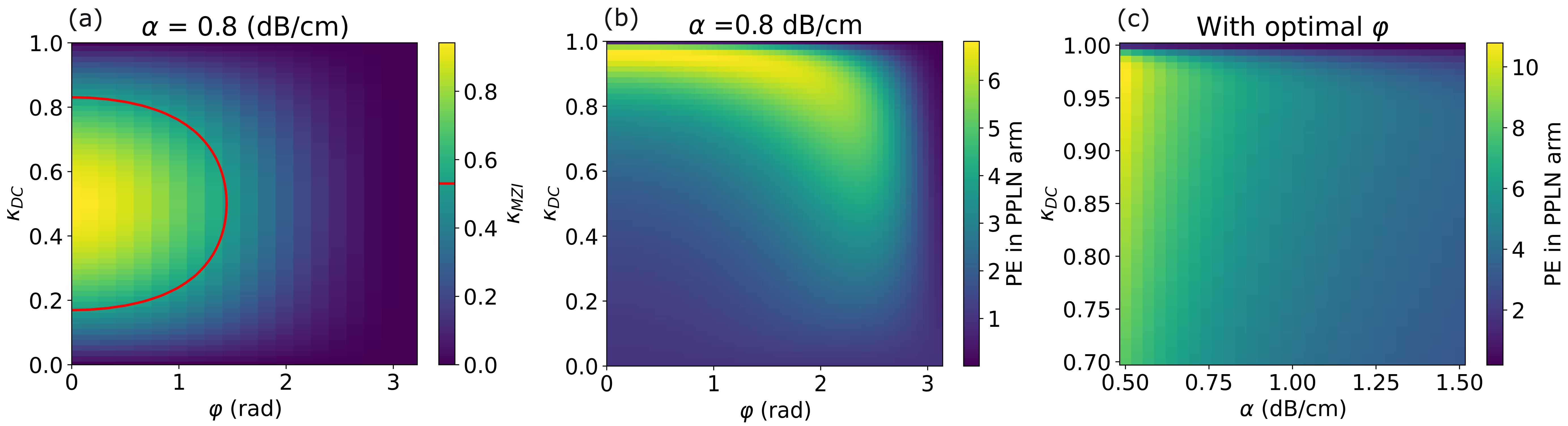}
\caption{(a) Simulation of the effective MZI coupling coefficient $\kappa_{\mathrm{MZI}}$ versus phase $\varphi$ and directional-coupler coupling $\kappa_{\mathrm{DC}}$ for a fixed propagation loss $\alpha = \SI{0.8}{\decibel\per\centi\meter}$. The red curve marks the experimentally measured maximum $\kappa_{\mathrm{MZI,max}}= 0.53$ (see Fig. \ref{fig:pannel1}(c)). Simulated pump power enhancement in the PPLN as a function of (b) the phase $\varphi$ and of $\kappa_{\mathrm{DC}}$ for $\alpha = \SI{0.8}{\decibel\per\centi\meter}$ and (c) of $\alpha$ and $\kappa_{\mathrm{DC}}$ at the phase $\varphi$ maximizing PE in the PPLN section.}
\label{fig:pannel2_a}
\end{figure}
In the simulation, $\varphi=0$ is the phase maximizing $\kappa_{\mathrm{MZI}}$ for a given $\kappa_{\mathrm{DC}}$ and therefore, two values of $\kappa_{\mathrm{DC}}$ for the tested sample are possible: 0.17 and 0.83. A cross section of Fig.\ref{fig:pannel2_a} (a) at $\varphi = 0$ is shown in the supplementary material in Fig. S3. Fig.~\ref{fig:pannel2_a} (b) shows the simulated pump power enhancement in the PPLN as a function of $\varphi$ and $\kappa_{\mathrm{DC}}$, for $\alpha = \SI{0.8}{\decibel\per\centi\meter}$. The results indicate that maximizing the pump enhancement requires a coupling coefficient $\kappa_{\mathrm{DC}} > 50\%$. Therefore, we can consider $\kappa_{\mathrm{DC}} = 0.83$ as the physically relevant solution, consistent with our design strategy of employing a high coupling coefficient to maximize PE. Fig.~\ref{fig:pannel2_a} (c) presents the PE in the PPLN section as a function of $\kappa_{\mathrm{DC}}$ and $\alpha$, evaluated at the optimal phase $\varphi$ ($\varphi=0$ in the SAX simulations). During the design phase, the exact post-fabrication values of $\alpha$ and $\kappa_{\mathrm{DC}}$ are inherently uncertain. Expecting propagation losses in the range $\SI{0.5}{\decibel\per\centi\meter} \leq \alpha \leq \SI{1.5}{\decibel\per\centi\meter}$ (including directional coupler losses), we intentionally selected $\kappa_{\mathrm{DC}}$ slightly below the nominal optimum (0.95 for $\alpha = \SI{0.8}{\decibel\per\centi\meter}$). This prevents operation in the regime where $\kappa_{\mathrm{DC}} \rightarrow 1$, in which critical coupling can no longer be achieved, resulting in a pronounced drop in PE, as can be seen in Fig.~\ref{fig:pannel2_a} (c). With this design choice, near-maximum PE in the PPLN arm can be recovered by tuning the MZI phase $\varphi$, as illustrated in Fig.~\ref{fig:pannel2_a} (b). This demonstrates a key advantage of the architecture: fabrication-induced variations in $\kappa_{\mathrm{DC}}$ and $\alpha$ can be compensated through phase tuning, enabling restoration of pump enhancement in the lower arm of the MZI.

\begin{figure}[H]
\centering\includegraphics[width=13.56 cm]{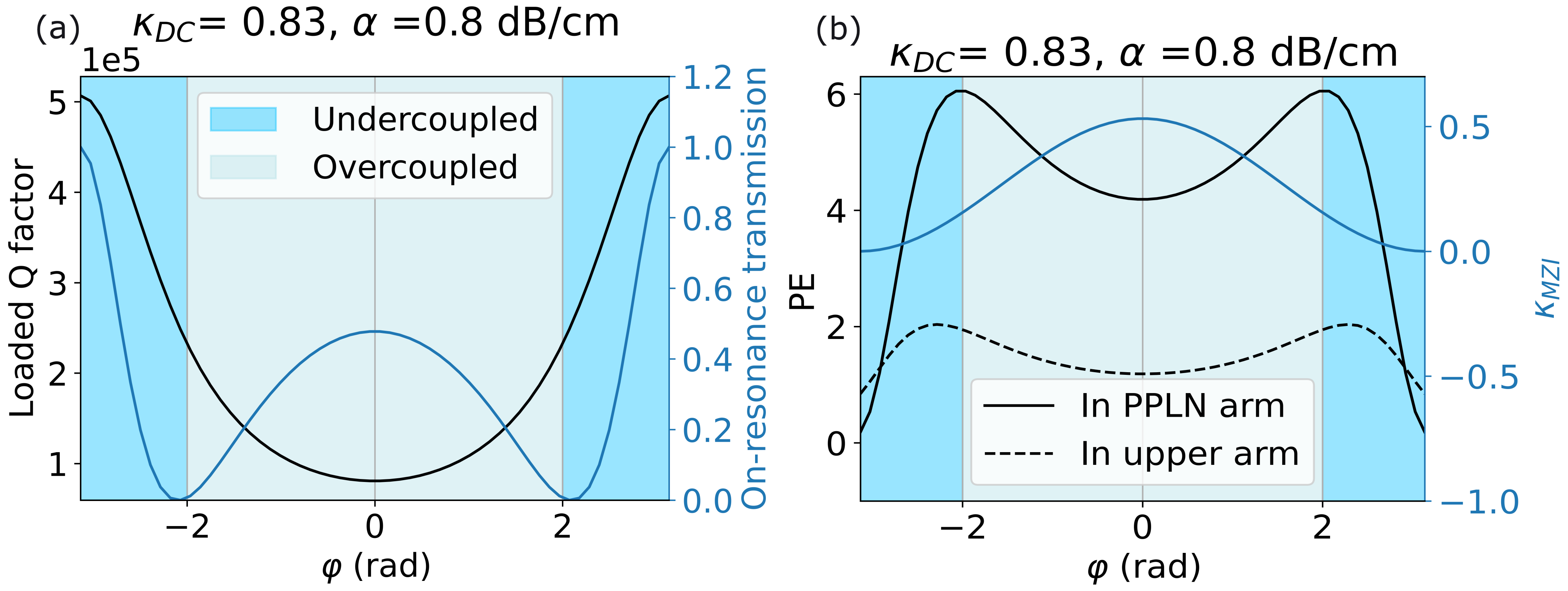}
\caption{(a) Simulated quality factor (left axis) and on-resonance transmission (right axis) as a function of $\varphi$. (b) Simulated pump power enhancement in the PPLN and upper MZI arm (left axis), together with $\kappa_{\mathrm{MZI}}$ (right axis), for $\kappa_{\mathrm{DC}}= 0.83$ and $\alpha = \SI{0.8}{\decibel\per\centi\meter}$. Colors indicate the under- and over-coupled regimes; at critical coupling, the pump enhancement in the PPLN is maximized.}
\label{fig:pannel2_b}
\end{figure}

We now simulate the full structure using $\alpha = \SI{0.8}{\decibel\per\centi\meter}$ and $\kappa_{\mathrm{DC}} = 0.83$. Fig.\ref{fig:pannel2_b} (a) shows the loaded quality factor (left axis), which well reproduces the experimental trend reported in Fig.\ref{fig:pannel1} (c), together with the on-resonance transmission (right axis). The latter allows identification of the transition between the under- and overcoupled regimes, as the on-resonance transmission is minimized at critical coupling. Fig.\ref{fig:pannel2_b} (b) shows the pump power enhancement in the lower arm of the MZI, where the PPLN is located, (black line, left axis) and in the upper arm (black dashed line, left axis). The PE in the PPLN arm is maximized at critical coupling, reaching $\mathrm{PE}=6$, while the enhancement in the upper MZI arm is $\mathrm{PE}=2$. Consequently, approximately \SI{25}{\percent} of the pump power does not contribute to frequency conversion. Using the optimal coupling value $\kappa_{\mathrm{DC}}=0.95$ would reduce this loss to \SI{15}{\percent}, but at the expense of reduced fabrication tolerance. When the pump is resonant, and since the SH field is not, the SH power scales with the square of the pump power according to
\begin{equation}
P_{\mathrm{SH}} \propto P_\mathrm{p}^2 = \mathrm{PE}^2 P_{\mathrm{p,in}}^2 ,
\label{eq:P_SH_vs_PE}
\end{equation}
where $P_{\mathrm{SH}}$ is the SH power, $P_\mathrm{p}$ is the pump power in the PPLN arm, and $P_{\mathrm{p,in}}$ is the launched pump power before injection into the racetrack. The pump power in the PPLN arm is maximized at critical coupling; accordingly, the SH output is also expected to peak in this regime. At critical coupling, $\mathrm{PE} = 6$, yielding an expected SH power enhancement factor of $\mathrm{SHE} = \mathrm{PE}^2 = 36$ compared to a straight PPLN waveguide of the same length, which doesn't provide resonant enhancement. Fig.\ref{fig:pannel2_b} (b) also shows $\kappa_{\mathrm{MZI}}$ as a function of $\varphi$ (right axis), whose trend closely reproduces the experimental results of Fig.\ref{fig:pannel1} (c).

\subsection{Nonlinear characterization}
\begin{figure}[H]
\centering\includegraphics[width=13.56 cm]{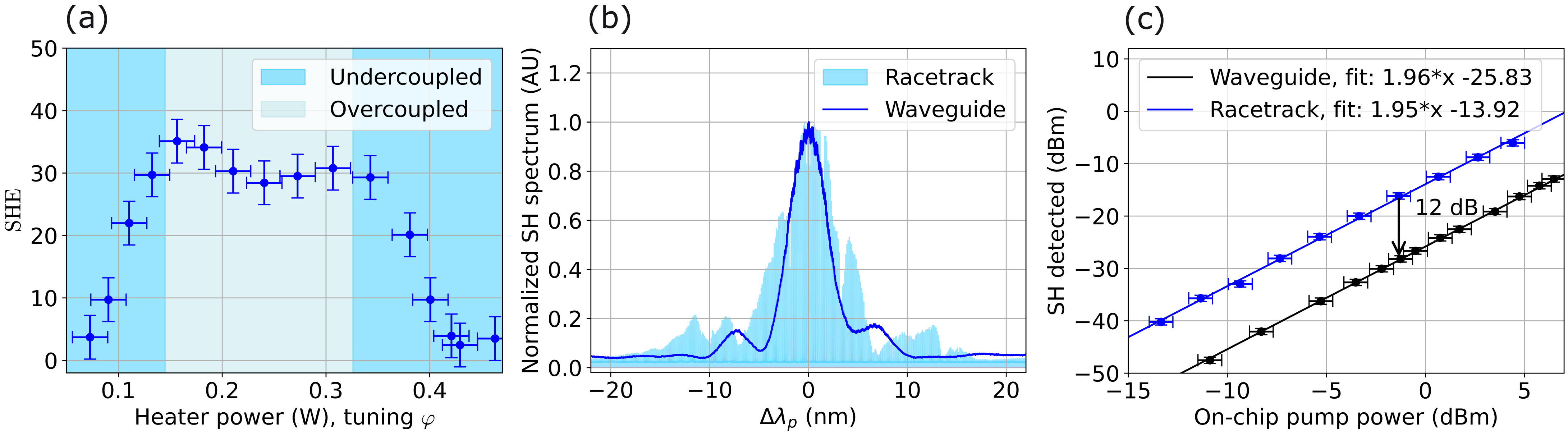}
\caption{(a) Experimental SH power enhancement as a function of the MZI phase $\varphi$, with maxima occurring at critical pump coupling (see Fig. \ref{fig:pannel2_b}(b)). (b) Experimental SH spectrum versus detuning from SH phase matching for the singly resonant racetrack (light blue) and a longer single-pass PPLN waveguide (dark blue). (c) SH power versus pump power showing a 12 dB enhancement for the racetrack (dark blue) compared to the single-pass PPLN (black). For the racetrack measurement, the pump is maintained on resonance and the phase $\varphi$ is set to achieve critical pump coupling. }
\label{fig:pannel3}
\end{figure}

We now present the experimental nonlinear characterization of the device. Fig.\ref{fig:pannel3} (a) shows the SH power enhancement as a function of MZI heater power, hence $\varphi$ and the coupling condition, while keeping the pump wavelength at resonance by adjusting the laser wavelength. The SH signal is measured using a photodetector, and its maximum is rescaled to the $\mathrm{SHE}$ value inferred in the following analysis. $\mathrm{SHE}$ is maximized when the pump is critically coupled, which agrees with the simulation in Fig.\ref{fig:pannel2_b} (b), where the pump power enhancement in the PPLN arm is also maximized at critical coupling. Note that the second maximum, observed at a heater power of 0.3 W, is slightly lower than the first. This effect can be attributed to residual thermal cross-talk: actuation of the MZI thermal heater induces a change in the PPLN temperature, which in turn alters its refractive index and phase-matching condition. As a result, the entire spectrum is shifted and its shape is distorted, leading to a reduction in peak height. 

Next, we compare the SHG conversion efficiency and bandwidth of the present device with those of a single-pass PPLN waveguide, the standard building block for frequency doubling in TFLN platforms. Both devices comprise short apodized sections at the input and output, and a longer central section with a lower chirp rate of the poling period. The period is chosen to have phase matching centered at 1550 nm. The single-pass PPLN has a total length of $L_1 = \SI{4400}{\micro\meter}$, including two apodization sections of $\SI{220}{\micro\meter}$ each, and a central chirp rate of $2\times10^{-6}$ $\mathrm{m/m}$. The racetrack-integrated PPLN has a total length of $L_2 = \SI{3250}{\micro\meter}$, apodization sections of $\SI{160}{\micro\meter}$ each, and a larger chirp rate of $9\times10^{-6}$ $\mathrm{m/m}$. Both waveguides share the same cross-section. The simulated normalized SHG conversion efficiency of an unchirped PPLN is $\SI{4000}{\%  \watt^{-1} \centi \meter ^{-2}}$. Taking into account the chirped poling period, the normalized peak efficiency is expected to decrease by a factor of 1.27 for the single-pass device and by a factor of 4 for the racetrack PPLN. Simulations predict a full width at half maximum (FWHM) SHG bandwidth, $\Delta \lambda$, of $\SI{22.3}{\nano\meter}$ for the racetrack PPLN and of $\SI{4.8}{\nano\meter}$ for the single-pass waveguide, corresponding to a 4.65-fold increase in bandwidth for the racetrack device. Experimentally, the SH spectrum shown in Fig.~\ref{fig:pannel3} (b) exhibits a FWHM bandwidth of $\SI{4.4}{\nano\meter}$ for the single-pass PPLN, in good agreement with simulations. For the racetrack, the measured bandwidth ranges between $\SI{6.5}{\nano\meter}$ and $\SI{7.9}{\nano\meter}$, depending on whether the FWHM is evaluated on the main peak or includes the adjacent lobe. Importantly, the SH spectrum of the racetrack PPLN is not continuous but consists of discrete peaks corresponding to the pump resonances. The SHG bandwidth in PPLNs is sensitive to dimensional inhomogeneities, such as thickness variations, which typically broaden the bandwidth while reducing the peak conversion efficiency. However, in the presence of a chirped poling period, the opposite behavior may arise if geometrical variations partially compensate the designed chirp. This mechanism likely explains why the experimentally observed bandwidth of the racetrack PPLN is smaller than predicted by simulations.

The power scaling of the SH as a function of the pump power, on-resonance and under critically coupled conditions, shown in Fig.~\ref{fig:pannel3} (c), demonstrates that the racetrack resonator provides a \SI{12}{\decibel} enhancement compared to the single-pass PPLN waveguide. As derived in section 4 of the supplementary information, the ratio of the SH peak powers scales as $\left( \frac{L_1}{L_2} \right) \left( \frac{L_{1,\mathrm{eff }}}{L_{2,\mathrm{eff}}} \right)$, where $L_i$ denotes the total length of the PPLN and $L_{i,\mathrm{eff}}$ the effective length of the interaction, that is the length of an ideal PPLN that would produce the measured bandwidth. Therefore, $L_{i,\mathrm{eff}}$ takes into account the combined effects of inhomogeneities and chirping. Since the phase-matching bandwidth is inversely proportional to the effective length, we can experimentally determine the ratio of effective lengths as $\frac{L_{1,\mathrm{eff}}}{L_{2,\mathrm{eff}}}= \frac{\Delta \lambda_2}{\Delta \lambda_1} = 1.63 \pm 0.16 $, where the uncertainty arises from the determination of $\Delta \lambda_2$. Rescaling the measured SH, to obtain a comparison of the efficiency independent on the device length, we obtain $\mathrm{SHE} = 10^{(12/10)} \left( \frac{L_1 }{L_2} \right) \left( \frac{\Delta \lambda_2}{\Delta \lambda_1} \right) = 35.1 \pm 3.5 $, in excellent agreement with the simulated value $\mathrm{SHE} = 36$.

\begin{figure}[htbp]
\centering\includegraphics[width=13.56 cm]{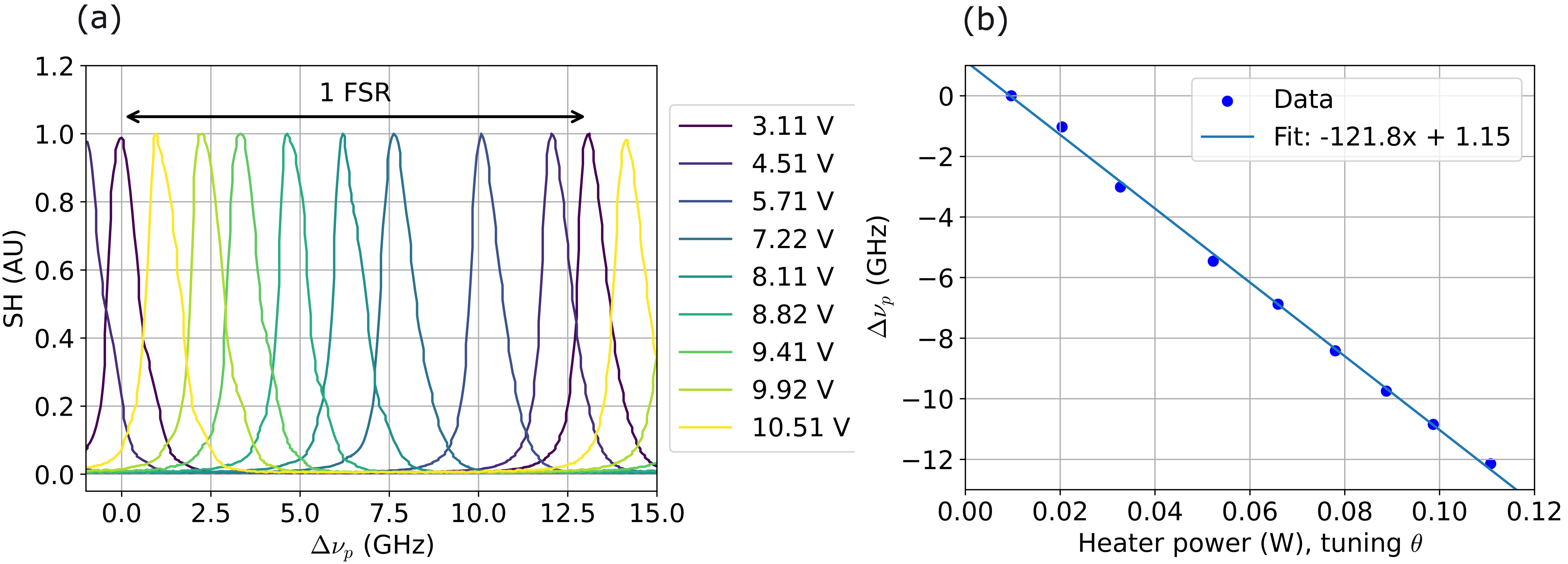} 
\caption{Continuous tuning of the SH peaks over one pump FSR. (a) SH peaks as a function of pump detuning from resonance for different applied voltages controlling the phase $\theta$. (b) Resonance peak shift as a function of heater power, which tunes $\theta$.  }
\label{fig:pannel4}
\end{figure}

We measure an on-chip normalized frequency conversion efficiency of $\eta_{\mathrm{1,SHG}} = \frac{P_{\mathrm{SH}}}{P_\mathrm{p}^2 L_1^2} = \SI{1300(100)}{\percent \per \watt \per \centi\meter\squared}$ for our PPLN waveguide, where $P_{\mathrm{SH}}$ is the detected off-chip SH power and $P_\mathrm{p}$ is the estimated on-chip pump power. The coupling loss at telecom wavelength is $\SI{11}{\decibel}$ per facet. This relatively high loss originates from the fact that the waveguides were simply tapered at the output, without employing double-etched inverse tapers, which are commonly used on the CSEM platform to achieve lower insertion losses \cite{sattari2024thin}. The racetrack exhibits an on-chip SHG efficiency of $\eta_{\mathrm{2,SHG}} = \frac{P_{\mathrm{SH}}}{P_\mathrm{p}^2} = \SI{4000(300)}{\percent \per \watt}$. Comparable efficiencies have been reported in \SI{15}{\milli\meter}-long PPLN waveguides with adapted poling \cite{chen2024adapted}, while the PPLN section used here is only \SI{3.25}{\milli\meter} long. The conversion efficiency of our PPLNs is limited by the waveguide's position between the positive and ground electrodes, as shown in the two-photon microscopy (TPM) images in Figs. S4 (a) and (b) in the supplementary, taken before and after waveguide patterning. Near the positive electrode, we observe near-complete inversion of the ferroelectric domains with a duty cycle of approximately $\SI{50}{\percent}$, while incomplete inversion occurs at the center and near the ground electrode. This issue can be mitigated by introducing a SiO$_2$ layer between the poling electrodes and the LN layer \cite{nagy2019reducing}, or by offsetting the waveguide position relative to the ground electrode, both of which could enhance efficiency. Note that the chirped poling period of the PPLN also tends to reduce the peak conversion efficiency. The SH spectrum in Fig.\ref{fig:pannel3} (b) shows a fine structure associated with the pump resonances, with a free spectral range (FSR) of 13.08 GHz. As shown in Figs.\ref{fig:pannel4} (a) and \ref{fig:pannel3} (b), tuning the heater within the racetrack, which controls the phase $\theta$, provides continuous tuning of the SH across the full pump FSR, resulting in a continuous SH spectrum with a $\sim$7 nm FWHM bandwidth. Fig. \ref{fig:pannel4} (b) yields a $V_{\pi}$ of \SI{5.18}{\volt} and a heater-induced pump tuning efficiency of \SI{122}{\giga\hertz\per\watt}.

\section{Discussion and conclusion}
In conclusion, we have demonstrated a singly resonant TFLN racetrack incorporating a tunable MZI coupler with an embedded short PPLN section. This architecture enables in situ control of the coupling regime while avoiding resonance alignment at widely separated wavelengths. The system can be actively tuned to critical coupling following the actual intrinsic losses of the cavity, maximizing in turn the pump power enhancement in the MZI arm containing the PPLN and thereby enhancing second harmonic generation. We achieve strong field enhancement in the PPLN, resulting in a second harmonic enhancement of 35 over a $\sim$7 nm bandwidth compared to an identical single-pass PPLN. Continuous tuning across one FSR at the fundamental harmonic is demonstrated, showing that, in principle, all the wavelengths included in the SHG spectrum can be accessed.
Although optimized here for SHG, this approach can be adapted to broadband nonlinear frequency conversion and quantum light generation. For example, in quantum applications requiring high escape efficiency in the telecom band, our design can be implemented by using $\kappa_{\mathrm{DC}} \sim 0.5$, as shown in Fig. \ref{fig:pannel2_b} (a). By changing the phase $\varphi$, one can then choose the best trade-off between escape efficiency and conversion efficiency.

\section{Back matter}

\begin{backmatter}
\bmsection{Funding}
Swiss National Science Foundation Bridge project 40B2-0 203480.

\bmsection{Acknowledgment}
This work is supported by SNSF Bridge project 40B2-0 203480.

\bmsection{Disclosures}
The authors declare no conflicts of interest.

\bmsection{Data availability} Data underlying the results presented in this paper are not publicly available at this time but may be obtained from the authors upon reasonable request.


\end{backmatter}

\bibliography{sample}

@article{kundu2025periodically,
  title={Periodically poled thin-film lithium niobate ring Mach Zehnder coupling interferometer as an efficient quantum light source},
  author={Kundu, Mrinmoy and Sikder, Bejoy and Huang, Heqing and Earnshaw, Mark and Al Sayem, Ayed},
  journal={Optics Express},
  volume={33},
  number={20},
  pages={43162--43175},
  year={2025},
  publisher={Optica Publishing Group}
}

@article{heine2026tunable,
  title={Tunable Coupler--Augmented Microrings: Reconfigurable Q-Factor Control for Foundry-Scale Quantum Light Sources},
  author={Heine, Jan and Dur{\'a}n G{\'o}mez, Juan Samuel Sebasti{\'a}n and Kues, Michael},
  journal={Nanophotonics},
  volume={15},
  number={6},
  pages={e70043},
  year={2026},
  publisher={Wiley Online Library}
}

@article{lu2019periodically,
  title={Periodically poled thin-film lithium niobate microring resonators with a second-harmonic generation efficiency of 250,000\%/W},
  author={Lu, Juanjuan and Surya, Joshua B and Liu, Xianwen and Bruch, Alexander W and Gong, Zheng and Xu, Yuntao and Tang, Hong X},
  journal={Optica},
  volume={6},
  number={12},
  pages={1455--1460},
  year={2019},
  publisher={OSA}
}

@inproceedings{sattari2024thin,
  title={Thin-film lithium niobate PICs: advancements and potential applications in telecom and beyond},
  author={Sattari, H and Prieto, I and Zarebidaki, H and Leo, J and Choong, G and Arefi, F and Orvietani, M and Della Torre, A and Mettraux, A and Dubois, F and others},
  booktitle={Integrated Photonics Platforms III},
  volume={13012},
  pages={86--89},
  year={2024},
  organization={SPIE}
}

@article{lu2021ultralow,
  title={Ultralow-threshold thin-film lithium niobate optical parametric oscillator},
  author={Lu, Juanjuan and Al Sayem, Ayed and Gong, Zheng and Surya, Joshua B and Zou, Chang-Ling and Tang, Hong X},
  journal={Optica},
  volume={8},
  number={4},
  pages={539--544},
  year={2021},
  publisher={Optical Society of America}
}

@article{briggs2024precise,
  title={Precise wavelength alignment of second-harmonic generation in thin-film lithium niobate resonators},
  author={Briggs, Ian and Chen, Paokang and Fan, Linran},
  journal={Optics Letters},
  volume={49},
  number={23},
  pages={6637--6640},
  year={2024},
  publisher={Optica Publishing Group}
}

@article{bruch2019chip,
  title={On-chip $\chi$\^{}(2) microring optical parametric oscillator},
  author={Bruch, Alexander W and Liu, Xianwen and Surya, Joshua B and Zou, Chang-Ling and Tang, Hong X},
  journal={Optica},
  volume={6},
  number={10},
  pages={1361--1366},
  year={2019},
  publisher={OSA}
}

@article{hwang2024spontaneous,
  title={Spontaneous parametric downconversion photon pair generation in small footprint X-cut periodically poled lithium niobate micro-resonator},
  author={Hwang, Hyeon and Noh, Woojin and Nurrahman, Mohamad Reza and Kim, Guhwan and Moon, Kiwon and Ju, Jung Jin and Lee, Hansuek and Seo, Min-Kyo},
  journal={Optics Letters},
  volume={49},
  number={19},
  pages={5379--5382},
  year={2024},
  publisher={Optica Publishing Group}
}

@article{chen2019ultra,
  title={Ultra-efficient frequency conversion in quasi-phase-matched lithium niobate microrings},
  author={Chen, Jia-Yang and Ma, Zhao-Hui and Sua, Yong Meng and Li, Zhan and Tang, Chao and Huang, Yu-Ping},
  journal={Optica},
  volume={6},
  number={9},
  pages={1244--1245},
  year={2019},
  publisher={OSA}
}

@article{mckenna2022ultra,
  title={Ultra-low-power second-order nonlinear optics on a chip},
  author={McKenna, Timothy P and Stokowski, Hubert S and Ansari, Vahid and Mishra, Jatadhari and Jankowski, Marc and Sarabalis, Christopher J and Herrmann, Jason F and Langrock, Carsten and Fejer, Martin M and Safavi-Naeini, Amir H},
  journal={Nature Communications},
  volume={13},
  number={1},
  pages={4532},
  year={2022},
  publisher={Nature Publishing Group UK London}
}

@article{ledezma2023octave,
  title={Octave-spanning tunable infrared parametric oscillators in nanophotonics},
  author={Ledezma, Luis and Roy, Arkadev and Costa, Luis and Sekine, Ryoto and Gray, Robert and Guo, Qiushi and Nehra, Rajveer and Briggs, Ryan M and Marandi, Alireza},
  journal={Science Advances},
  volume={9},
  number={30},
  pages={eadf9711},
  year={2023},
  publisher={American Association for the Advancement of Science}
}

@article{roy2023visible,
  title={Visible-to-mid-IR tunable frequency comb in nanophotonics},
  author={Roy, Arkadev and Ledezma, Luis and Costa, Luis and Gray, Robert and Sekine, Ryoto and Guo, Qiushi and Liu, Mingchen and Briggs, Ryan M and Marandi, Alireza},
  journal={Nature Communications},
  volume={14},
  number={1},
  pages={6549},
  year={2023},
  publisher={Nature Publishing Group UK London}
}

@article{park2024single,
  title={Single-mode squeezed-light generation and tomography with an integrated optical parametric oscillator},
  author={Park, Taewon and Stokowski, Hubert and Ansari, Vahid and Gyger, Samuel and Multani, Kevin KS and Celik, Oguz Tolga and Hwang, Alexander Y and Dean, Devin J and Mayor, Felix and McKenna, Timothy P and others},
  journal={Science Advances},
  volume={10},
  number={11},
  pages={eadl1814},
  year={2024},
  publisher={American Association for the Advancement of Science}
}

@article{nagy2019reducing,
  title={Reducing leakage current during periodic poling of ion-sliced x-cut MgO doped lithium niobate thin films},
  author={Nagy, Jonathan Tyler and Reano, Ronald M},
  journal={Optical Materials Express},
  volume={9},
  number={7},
  pages={3146--3155},
  year={2019},
  publisher={OSA}
}

@article{chen2024adapted,
  title={Adapted poling to break the nonlinear efficiency limit in nanophotonic lithium niobate waveguides},
  author={Chen, Pao-Kang and Briggs, Ian and Cui, Chaohan and Zhang, Liang and Shah, Manav and Fan, Linran},
  journal={Nature Nanotechnology},
  volume={19},
  number={1},
  pages={44--50},
  year={2024},
  publisher={Nature Publishing Group UK London}
}

@misc{csem_foundry,
    author = {CSEM},
    title = {Thin-Film Lithium Niobate Foundry Services by CSEM},
    year = 2025,
    howpublished = {\url{https://www.csem.ch/en/tailored-services/tfln-foundry-services/}},
    note = {Accessed: 2025-04-07}
}

@article{clementi2025ultrabroadband,
  title={Ultrabroadband milliwatt-level resonant frequency doubling on a chip},
  author={Clementi, Marco and Zatti, Luca and Zhou, Ji and Liscidini, Marco and Br{\`e}s, Camille-Sophie},
  journal={Nature Communications},
  volume={16},
  number={1},
  pages={6164},
  year={2025},
  publisher={Nature Publishing Group UK London}
}

@article{lukin20204h,
  title={4H-silicon-carbide-on-insulator for integrated quantum and nonlinear photonics},
  author={Lukin, Daniil M and Dory, Constantin and Guidry, Melissa A and Yang, Ki Youl and Mishra, Sattwik Deb and Trivedi, Rahul and Radulaski, Marina and Sun, Shuo and Vercruysse, Dries and Ahn, Geun Ho and others},
  journal={Nature Photonics},
  volume={14},
  number={5},
  pages={330--334},
  year={2020},
  publisher={Nature Publishing Group UK London}
}

@article{lu2021efficient,
  title={Efficient photoinduced second-harmonic generation in silicon nitride photonics},
  author={Lu, Xiyuan and Moille, Gregory and Rao, Ashutosh and Westly, Daron A and Srinivasan, Kartik},
  journal={Nature Photonics},
  volume={15},
  number={2},
  pages={131--136},
  year={2021},
  publisher={Nature Publishing Group UK London}
}

@article{nitiss2022optically,
  title={Optically reconfigurable quasi-phase-matching in silicon nitride microresonators},
  author={Nitiss, Edgars and Hu, Jianqi and Stroganov, Anton and Br{\`e}s, Camille-Sophie},
  journal={Nature Photonics},
  volume={16},
  number={2},
  pages={134--141},
  year={2022},
  publisher={Nature Publishing Group UK London}
}

@article{lake2016efficient,
  title={Efficient telecom to visible wavelength conversion in doubly resonant gallium phosphide microdisks},
  author={Lake, David P and Mitchell, Matthew and Jayakumar, Harishankar and Dos Santos, La{\'\i}s Fujii and Curic, Davor and Barclay, Paul E},
  journal={Applied Physics Letters},
  volume={108},
  number={3},
  year={2016},
  publisher={AIP Publishing}
}

@article{bruch201817,
  title={17 000\%/W second-harmonic conversion efficiency in single-crystalline aluminum nitride microresonators},
  author={Bruch, Alexander W and Liu, Xianwen and Guo, Xiang and Surya, Joshua B and Gong, Zheng and Zhang, Liang and Wang, Junxi and Yan, Jianchang and Tang, Hong X},
  journal={Applied Physics Letters},
  volume={113},
  number={13},
  year={2018},
  publisher={AIP Publishing}
}

@article{desiatov2019ultra,
  title={Ultra-low-loss integrated visible photonics using thin-film lithium niobate},
  author={Desiatov, Boris and Shams-Ansari, Amirhassan and Zhang, Mian and Wang, Cheng and Lon{\v{c}}ar, Marko},
  journal={Optica},
  volume={6},
  number={3},
  pages={380--384},
  year={2019},
  publisher={Optical Society of America}
}

@article{stefano2024broadband,
  title={Broadband spontaneous parametric downconversion in reconfigurable poled linearly uncoupled resonators},
  author={Stefano, Alessia and Zatti, Luca and Liscidini, Marco},
  journal={Optics Letters},
  volume={49},
  number={17},
  pages={4819--4822},
  year={2024},
  publisher={Optica Publishing Group}
}

@article{wang2018ultrahigh,
  title={Ultrahigh-efficiency wavelength conversion in nanophotonic periodically poled lithium niobate waveguides},
  author={Wang, Cheng and Langrock, Carsten and Marandi, Alireza and Jankowski, Marc and Zhang, Mian and Desiatov, Boris and Fejer, Martin M and Lon{\v{c}}ar, Marko},
  journal={Optica},
  volume={5},
  number={11},
  pages={1438--1441},
  year={2018},
  publisher={Optical Society of America}
}

@article{yariv2000universal,
  title={Universal relations for coupling of optical power between microresonators and dielectric waveguides},
  author={Yariv, Amnon},
  journal={Electronics letters},
  volume={36},
  number={4},
  pages={321--322},
  year={2000},
  publisher={IET}
}

@article{phillips2010efficiency,
  title={Efficiency and phase of optical parametric amplification in chirped quasi-phase-matched gratings},
  author={Phillips, CR and Fejer, MM},
  journal={Optics letters},
  volume={35},
  number={18},
  pages={3093--3095},
  year={2010},
  publisher={Optical Society of America}
}

\clearpage
\appendix
\setcounter{page}{1}

\section*{Supplementary Information to: Enhancing  broadband second harmonic generation in a thin film lithium niobate racetrack resonator with tunable-coupling}
\addcontentsline{toc}{section}{Supplementary Information}

\renewcommand{\thefigure}{S\arabic{figure}}
\renewcommand{\theequation}{S\arabic{equation}}
\renewcommand{\thetable}{S\arabic{table}}

\setcounter{figure}{0}
\setcounter{equation}{0}
\setcounter{table}{0}

\section{Derivation of the pump field enhancement in the PPLN section}

\begin{figure}[H]
\centering\includegraphics[width=11 cm]{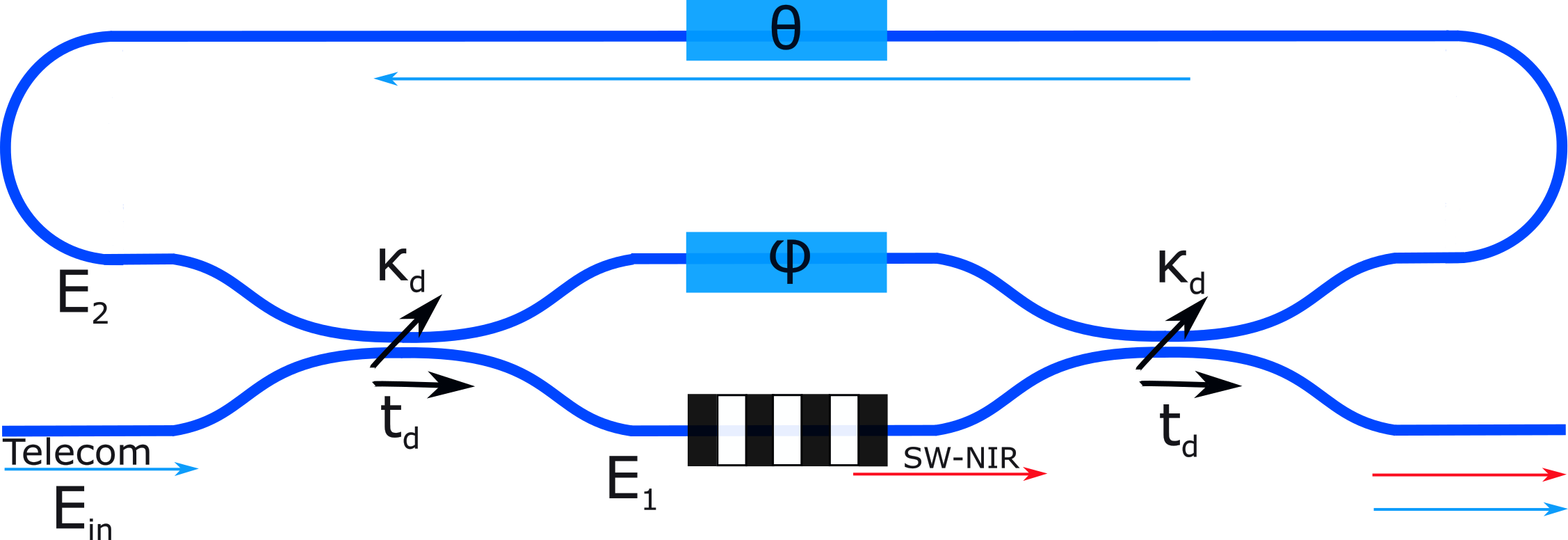}
\caption{Doubly resonant racetrack with a MZI coupler and a PPLN in its lower arm.}
\label{S1}
\end{figure}

In this section we derive an analytical solution of the pump electrical field at the input of the PPLN, denoted as $E_1$ in Fig. \ref{S1}, from which we infer the pump field enhancement 
\begin{equation}
\mathrm{FE}= \frac{E_1}{E_{in}},
\end{equation}
where $E_{in}$ is the electric field at the input. The field coupling and transmission coefficients of the directional coupler are expressed as
\begin{align}
\kappa_d &= -j\sqrt{\kappa_{\mathrm{DC}}} \\
t_d &= \sqrt{1-\kappa_d^2}.
\end{align}
Note that in the main text we use the power coupling coefficient $\kappa_{\mathrm{DC}} = |\kappa_d|^2 $. The length of the racetrack (not including the MZI section) is denoted by $L_r$ and the arms of the MZI are approximated by the PPLN length, $L$. We consider propagation losses, $\alpha$, to occur in the sections $L_r$ and $L$ and neglect them in the directional couplers. 
At all time, the electric field in front of the PPLN can be expressed as 
\begin{equation}
E_1 = E_{in}t_d + E_2\kappa_d.
\label{Eq1}
\end{equation}
The location of $E_2$ is indicated in Fig. \ref{S1}. In what follows, we want to derive an expression for $E_2$ after each round trip. After 1 round trip, we have
\begin{equation}
    E^{(1)}_2 = E_{in}t_d e^{-\frac{\alpha}{2} L}e^{i\beta L}\kappa_d e^{-\frac{\alpha}{2}L_r}e^{i\beta L_r} + 
     E_{in} \kappa_d  e^{-\frac{\alpha}{2} L}e^{i\beta L} e^{i\varphi}t_d e^{-\frac{\alpha}{2}L_r}e^{i\beta L_r},
\end{equation}
where $\beta$ is the pump wavector. After 2 round trips, the field becomes
\begin{equation}
\begin{split}
    E^{(2)}_2 & = E^{(1)}_2\Big[ 1 + \kappa^2_d e^{-\frac{\alpha}{2} (L+L_r)}e^{i\beta(L+L_r)} +  
    t^2_d e^{i\varphi}e^{-\frac{\alpha}{2} (L+L_r)}e^{i\beta(L+L_r)}  \Big] \\
    &:= E^{(1)}_2\Big[ 1 + A + B \Big].
    \end{split}
\end{equation}
We can rewrite it as 
\begin{equation}
    E^{(2)}_2 = E^{(1)}_2 + E^{(1)}_2\Big[A + B \Big] = E^{(1)}_2\Bigg[ 1 + \Big[A + B \Big]\Bigg].
\end{equation}
After 3 round trips, the field is equal to
\begin{equation}
    E^{(3)}_2 = E^{(1)}_2 + E^{(2)}_2\Big[ A + B \Big] = E^{(1)}_2\Bigg[ 1 + \Big[A + B \Big] + \Big[A + B \Big]^2\Bigg].
\end{equation}
Finally, after n round trips we have
\begin{equation}
    E^{(n)}_2 = E^{(1)}_2 + E^{(n-1)}_2\Big[ A + B \Big] =  E^{(1)}_2\Bigg[ 1 +\Big[ A + B\Big] + ... +\Big[ A + B\Big]^{n-1} \Bigg],
\end{equation}
which can be expressed as a finite geometric serie
\begin{equation}
    E^{(n)}_2 =  E^{(1)}_2\sum_{k=0}^{n-1} \Big[ A + B\Big]^k.
    \label{Eq7}
\end{equation}
The finite geometric serie 
\begin{equation}
    S_n = a + ar + ar^2 + ar^3 + ... + ar^{n-1 }= \sum^{n-1}_{k=0}  a r^k= 
\begin{cases}
a\frac{1-r^n}{1-r} \quad \text{if }r\neq 1 \\
na , \quad \quad   \text{if }r = 1 \\
\end{cases}
\end{equation}
also has a solution when $n\rightarrow \infty$, providing $|r|< 1$ to ensure convergence:
\begin{equation}
 S_{n \rightarrow \infty} = \sum_{k=0}^{n\rightarrow \infty}  a r^k = \frac{a}{1-r}. 
\end{equation}
We can now rewrite Eq. \ref{Eq7} as
\begin{equation}
E^{(n\rightarrow\infty)}_2 = \frac{E^{(1)}_2 }{1-A-B} = \frac{E_{in}t_d\kappa_d e^{-\frac{\alpha}{2} (L+L_r)}e^{i\beta(L+L_r)}\Big[ 1+ e^{i\varphi}\Big]}{1- e^{-\frac{\alpha}{2} (L+L_r)}e^{i\beta(L+L_r)}\Big[ k_d^2 + t_d^2 e^{i\varphi}\Big]  }.
\label{Eq_sol}
\end{equation}
Note that Eq.\ref{Eq_sol} is only valid if $|A+B| < 1$, which is the case as $|\kappa_d|^2 + |t_d|^2  = 1$. Indeed, the serie would only diverge in the case of lossless propagation for which $ \alpha = 0 $. Thus, for realistic case, the serie converges. Using Eqs. \ref{Eq1} and  \ref{Eq_sol}, we can now express the electric field in the lower arm of the MZI (at the input of the PPLN) as
\begin{equation}
    E_1 = E_{in}t_d + E^{(n\rightarrow\infty)}_2\kappa_d =  E_{in}\Bigg[t_d + \frac{  \kappa_d^2 t_d e^{-\frac{\alpha}{2} (L + L_r)}e^{i\beta(L + L_r)}\Big[ 1 + e^{i\varphi}\Big]}{1-e^{-\frac{\alpha}{2} (L + L_r)}e^{i\beta(L + L_r)}\Big[\kappa_d^2 + t_d^2 e^{i\varphi}\Big]}  \Bigg]. 
\end{equation}
and the field enhancement factor is given by:
\begin{equation}
\mathrm{FE}= \frac{E_1}{E_{in}} = \Bigg[t_d + \frac{  \kappa_d^2 t_d e^{-\frac{\alpha}{2} (L + L_r)}e^{i\beta(L + L_r)}\Big[ 1 + e^{i\varphi}\Big]}{1-A-B}  \Bigg].
\label{Eq:FE_derivation}
\end{equation}

\newpage

\section{Comparison of the SAX model with the analytical derivation}
\begin{figure}[H]
\centering\includegraphics[width=7 cm]{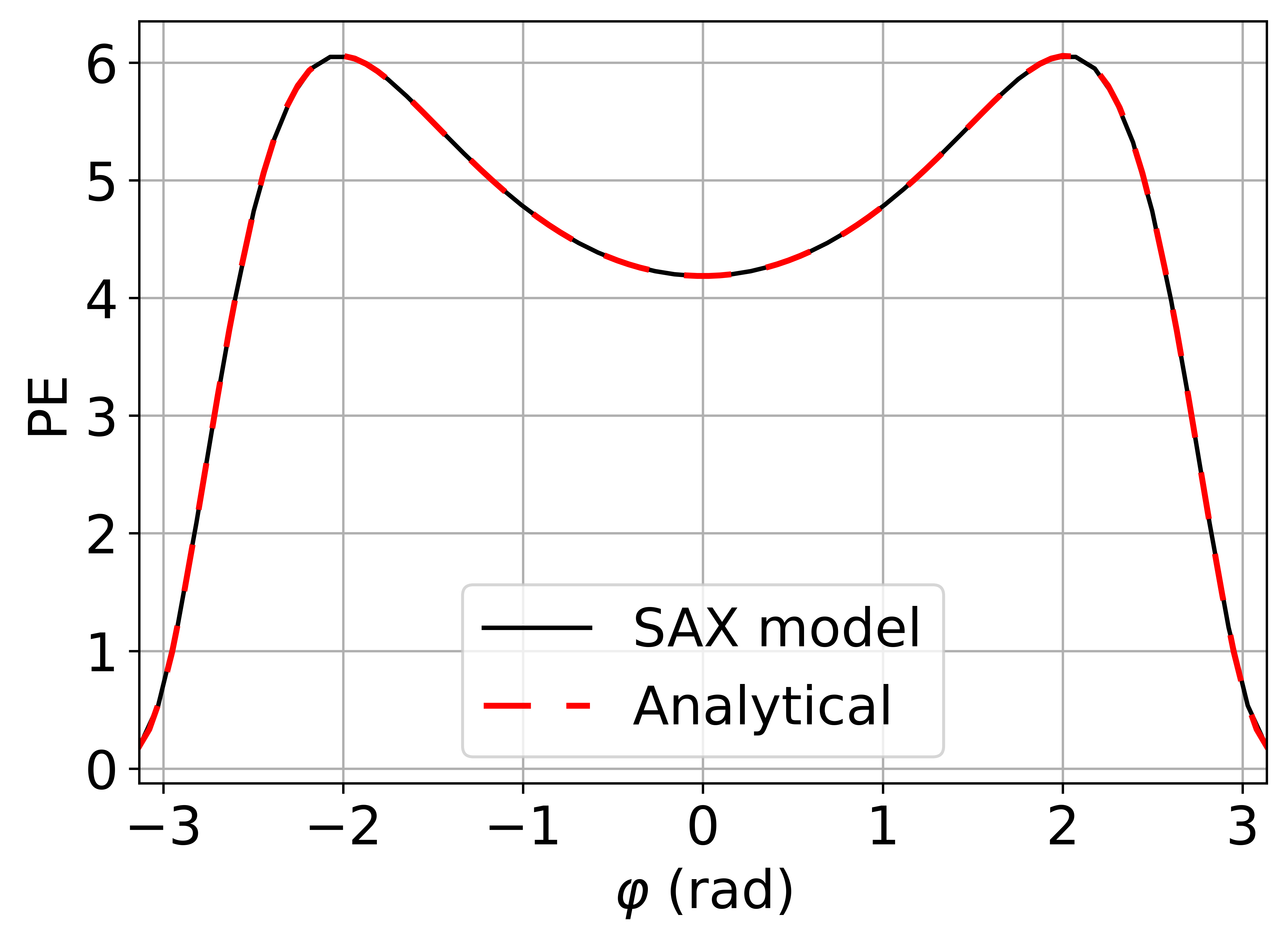}
\caption{PE as a function of $\varphi$, obtained with the SAX model and with the analytical formula in Eq. \ref{Eq:FE_derivation}, for $\kappa_{\mathrm{DC}}= 0.83$ and $\alpha = \SI{0.8}{\decibel \per \centi \meter}$.}
\label{S2}
\end{figure}

We now compare the simulations from the SAX model with the analytical formula in Eq. \ref{Eq:FE_derivation} (taking PE = |FE|$^2$), that we derived in the previous section. Fig. \ref{S2} shows that they perfectly overlap. 

\section{Simulations}
\begin{figure}[H]
\centering\includegraphics[width=7 cm]{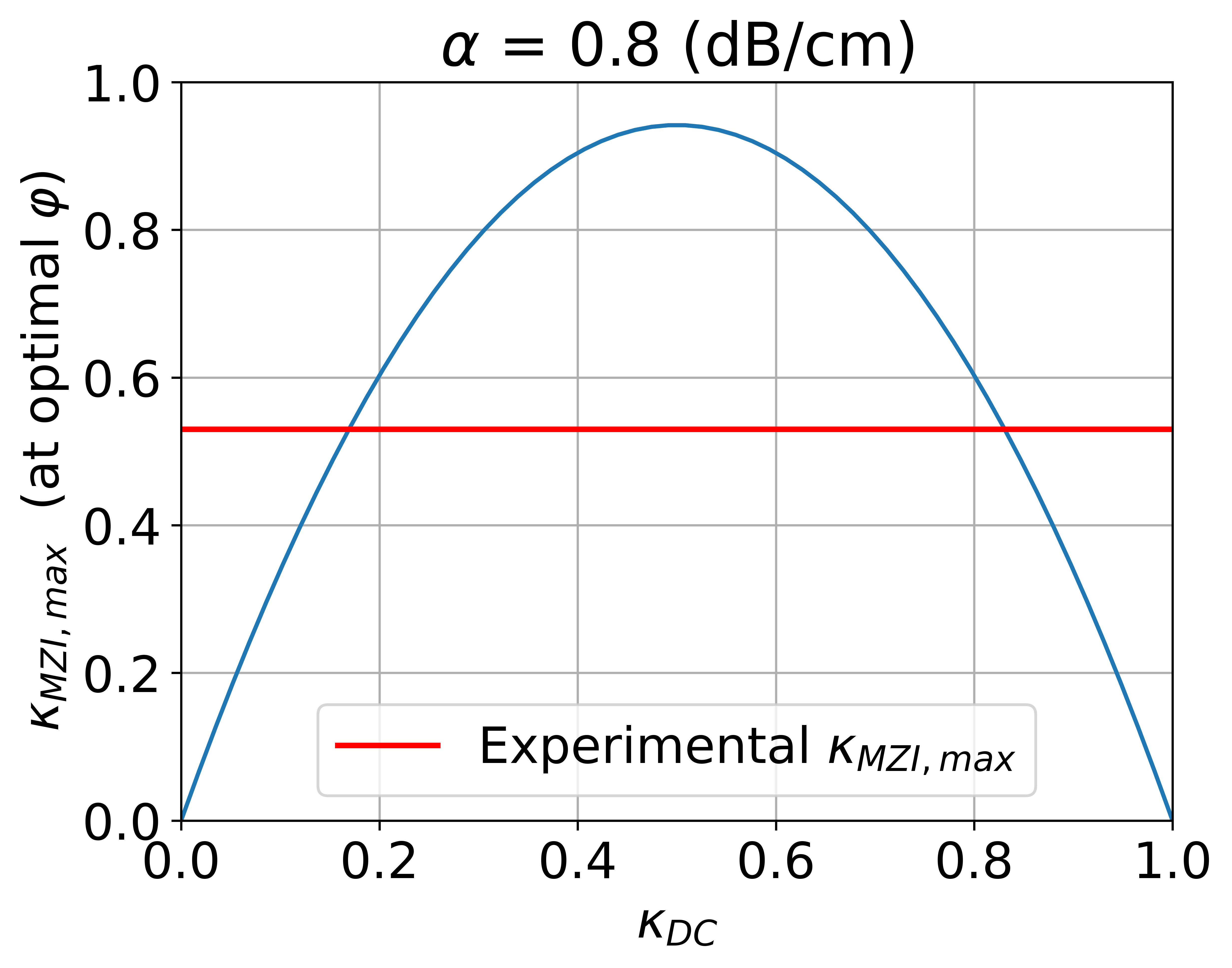}
\caption{Simulated maximum $\kappa_{\mathrm{MZI}}$ (at optimal $\varphi$) as a function of $\kappa_{\mathrm{DC}}$. Its intersection with the experimental value (red) yields two possible solutions; using the design value, the physical solution is identified as $\kappa_{\mathrm{DC}}= 0.83$.}
\label{fig:pannelS3}
\end{figure}

From the maximal coupling coefficient, $\kappa_{\mathrm{DC, max}}$, and the propagation loss measured experimentally, we infer the coupling coefficient of the directional coupler, $\kappa_{\mathrm{DC}}$. Fig. \ref{fig:pannelS3} shows a cross section of Fig. 2 (a) in the main text, at the phase maximizing $\kappa_{\mathrm{MZI}}$ ($\varphi = 0$ in the simulations). The intersections between the experimental value $\kappa_{\mathrm{MZI,max}}$ (red line) and the simulated $\kappa_{\mathrm{MZI}}(\kappa_{\mathrm{DC}})$ (blue curve) identify the two possible values of $\kappa_{\mathrm{DC}}$. The physically relevant solution is $\kappa_{\mathrm{DC}} = 0.83$, consistent with the design choice of a high coupling coefficient to maximize the pump field enhancement.

\section{Derivation of the SH peak power ratio between PPLNs with different chirp and inhomogeneities}

In this section, we derive the ratio of the SH peak powers for PPLNs whose SHG bandwidth is intentionally and/or un-intentionally increased, for example by geometrical inhomogeneities and/or by a chirped poling period. We begin by considering two ideal PPLNs with perfectly homogeneous cross-sections and no chirp. Their lengths, SH spectral areas, bandwidths, and peak powers are denoted by $L_i$, $A_i$, $\Delta \lambda_i$, and $P_{\mathrm{peak},i}$, respectively, with $i=1,2$. Assuming $L_1>L_2$, it follows that $A_1 > A_2$, $P_{\mathrm{peak},1}> P_{\mathrm{peak},2}$ and $\Delta \lambda_1 <\Delta \lambda_2$. The SH spectral area scales with the product of the peak power and the bandwidth:
\begin{equation}
A_i \propto P_{\mathrm{peak},i}\Delta\lambda_i.
\label{Equ_2}
\end{equation}
Under a simplified rectangular approximation of the spectrum, the area reduces to the peak power multiplied by the bandwidth, $A_i =P_{\mathrm{peak},i}\Delta\lambda_i$. This approximation does not affect the validity of the derivation below, but merely simplifies the analytical treatment. The bandwidth scales inversely with the device length,
\begin{equation}
\Delta\lambda_i \propto \frac{1}{L_i},
\label{Equ_4}
\end{equation}
and, in the undepleted regime, the SH peak power scales quadratically with length,
\begin{equation}
P_{\mathrm{peak},i} \propto L_i^2 .
\label{Equ_3}
\end{equation}
We now consider PPLNs of identical physical length that are affected by different geometrical inhomogeneities and chirped poling period. We indicate the actual SH peak power and bandwidth as $P_{\mathrm{peak},i}$ and $\Delta\lambda_i'$ with  integrated SH spectral area $A^{'}_i$ given by:
\begin{equation}
A_i' \propto P_{\mathrm{peak},i}'\Delta\lambda_i' .
\label{Equ_1}
\end{equation}
Chirping redistributes the phase matching spectrum and conversion efficiency over a wider bandwidth \cite{phillips2010efficiency}. Effectively, the integrated SH spectral area is conserved with respect to an ideal, non-chirped waveguide, so that $A^{'}_i= A_i$.

To account for the effect of chirp and inhomogeneities, we define an effective interaction length from the measured bandwidth as:
\begin{equation}
L_{i,\mathrm{eff}} \propto \frac{1}{\Delta\lambda_i'} .
\label{Equ_6}
\end{equation}
We are interested in the ratio $\frac{P^{'}_{\mathrm{peak},2}}{P^{'}_{\mathrm{peak},1}}$. From Eq.~\ref{Equ_1}, we obtain
\begin{equation}
\frac{P_{\mathrm{peak},2}'}{P_{\mathrm{peak},1}'} \propto
\frac{A_2}{\Delta\lambda_2'}
\frac{\Delta\lambda_1'}{A_1}.
\end{equation}
Using Eq.~\ref{Equ_2}, this becomes
\begin{equation}
\frac{P_{\mathrm{peak},2}'}{P_{\mathrm{peak},1}'} \propto 
\frac{P_{\mathrm{peak},2},\Delta\lambda_2}
{P_{\mathrm{peak},1},\Delta\lambda_1}
\frac{\Delta\lambda_1'}{\Delta\lambda_2'} .
\end{equation}
Finally, combining Eqs.~\ref{Equ_4} and \ref{Equ_3}, we obtain
\begin{equation}
\frac{P_{\mathrm{peak},2}'}{P_{\mathrm{peak},1}'} \propto
\frac{L_2}{L_1}
\frac{\Delta\lambda_1'}{\Delta\lambda_2'} .
\label{Equ_5_useful_one}
\end{equation}
Importantly, we can experimentally determine the ratio $\frac{\Delta \lambda^{'}_1}{\Delta \lambda^{'}_2}$ from the SHG spectra of the two waveguides. In the main text, Eq.~\ref{Equ_5_useful_one} is employed to rescale the measured SH peak powers, thereby isolating the efficiency enhancement originating solely from the resonant structure itself, independent of differences in PPLN length, chirp rate, and fabrication-induced inhomogeneities. For this reason, the experimental data are divided by the ratio given in Eq.~\ref{Equ_5_useful_one}. Substituting Eq.~\ref{Equ_6} into Eq.~\ref{Equ_5_useful_one}, we recover the familiar quadratic scaling of SH power with interaction length:
\begin{equation}
\frac{P_{\mathrm{peak},2}'}{P_{\mathrm{peak},1}'} \propto 
\frac{L_2}{L_1}
\frac{L_{2,\mathrm{eff}}}{L_{1,\mathrm{eff}}}.
\end{equation}

\section{Two-photon microscopy images}

\begin{figure}[H]
\centering\includegraphics[width=9 cm]{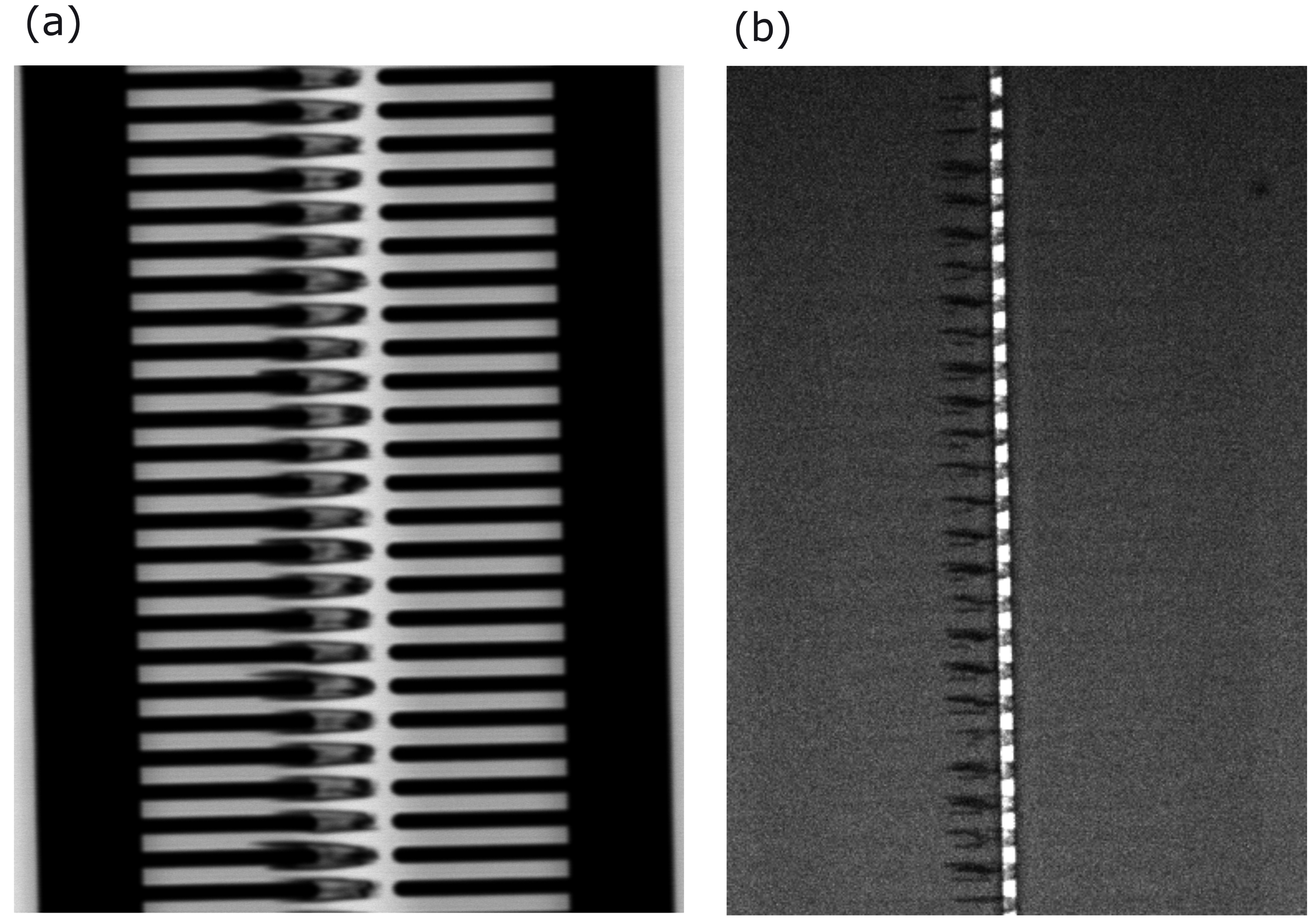}
\caption{Two-photon microscopy images taken (a) before etching and (b) after etching the waveguides. The distance between the tips of the electrodes is \SI{4.28}{\micro \meter}. The positive electrode is on the left side of the image, while the ground electrode is on the right.}
\label{fig:pannelS5}
\end{figure}

Figs. \ref{fig:pannelS5} (a) and (b) show two-photon microscopy (TPM) images taken before and after waveguide patterning. In fig. \ref{fig:pannelS5} (a), near the positive electrode, we observe near-complete inversion of the ferroelectric domains with a duty cycle of approximately $\SI{50}{\percent}$, while incomplete inversion occurs at the center and near the ground electrode. This issue can be mitigated by introducing a SiO$_2$ layer between the poling electrodes and the LN layer \cite{nagy2019reducing}. As the waveguide are centered between the positive and ground electrodes, the poling is thus incomplete in the PPLN cross-section, which limits the conversion efficiency. Offsetting the waveguide position relative to the ground electrode or adding a SiO$_2$ layer between the electrodes and the LN layer, could improve the conversion efficiency.

\end{document}